\documentclass[
    ,final            
,12pt
  ]
  {aipproc}

\layoutstyle{8x11single}
\usepackage{latexsym,epsfig,amsmath,amssymb}
\usepackage{amssymb}
\usepackage{amsmath}
\usepackage{amsfonts}
\usepackage{epsfig} 
\usepackage{verbatim} 
\usepackage{setspace}
\usepackage{graphics, graphicx}
\newcommand{\seq}{\begin{subequations}}
\newcommand{\sen}{\end{subequations}}
\newcommand{\eq}{\begin{eqnarray}}
\newcommand{\en}{\end{eqnarray}}
\newcommand{\ra}{\rangle}

\begin{document}

\title{Possible hadronic molecule structure of\\the Y(3940) and Y(4140)}

\classification{12.38.Lg, 12.39.Fe, 13.25.Jx, 14.40.Gx}
\keywords      {charm mesons, hadronic molecule, strong and radiative decay}

\author{Tanja Branz\footnote{tanja.branz@uni-tuebingen.de}}{
  address={Institut f\"ur Theoretische Physik, Universit\"at T\"ubingen,\\
       Auf der Morgenstelle 14, D$-$72076 T\"ubingen, Germany
         }
}

\author{Thomas Gutsche}{
}

\author{Valery E. Lyubovitskij\footnote{On leave of absence from Department of Physics, Tomsk State
University, 634050 Tomsk, Russia}}{
}

\begin{abstract}
In the present article we report on evidence for hadronic molecule structures in the charmonium sector. In particular we discuss the 
$Y(3940)$ and the recently observed $Y(4140)$ as heavy hadron molecule states 
with quantum numbers $J^{\rm PC} = 0^{++}$ or $2^{++}$. The $Y(3940)$ state is
considered to be a superposition of 
$D^{\ast +}D^{\ast -}$ and $D^{\ast 0} \overline{D^{\ast 0}}$, 
while the $Y(4140)$ is a bound state of $D_s^{\ast +}$ and 
$D_s^{\ast -}$ mesons. We give predictions for both the strong
$Y(3940) \to J/\psi \omega$, $Y(4140) \to J/\psi \phi$
and radiative $Y(3940)/Y(4140) \to \gamma\gamma$ decay widths in
a phenomenological Lagrangian approach.
The results for the strong hidden charm decay modes
clearly support the molecular interpretation of the $Y(3940)$
and $Y(4140)$, while our estimates for the radiative decays provide a sensitive test for the underlying meson structure of the two $Y$ mesons discussed here. The alternative assignment of
$J^{\rm PC} = 2^{++}$ is also tested, giving similar results for
the strong decay widths.

\end{abstract}

\maketitle

\section{Introduction}

With the discovery of the charmonium-like $X$, $Y$ and $Z$ mesons the prospects were raised of possibly identifying a non-trivial structure of a meson resonance. The latest announcement \cite{Aaltonen:2009tz} of the narrow
state $Y(4140)$ by the CDF Collaboration at Fermilab is just the continuation of a series of previously discovered charmonium-like resonances which are not easily interpreted as conventional quark-antiquark structures. 
Now the CDF Collaboration has evidence 
of a narrow near-threshold structure, termed the $Y(4140)$ meson, 
in the $J/\psi \phi$ mass spectrum in exclusive $B^+ \to J/\psi \phi K^+$
decays with the mass $m_{Y(4140)} = 4143.0 \pm 2.9 ({\rm stat}) \pm 
1.2 ({\rm syst})$ MeV and natural width $\Gamma_{Y(4140)} = 11.7^{+8.3}_{-5.0} ({\rm stat}) 
\pm 3.7 ({\rm syst})$ MeV~\cite{Aaltonen:2009tz}.
As already stressed in \cite{Aaltonen:2009tz}, the new structure $Y(4140)$,
which decays to $J/\psi \phi$ just above the $J/\psi \phi$ threshold, 
shows a similar decay and production pattern to the previously discovered 
$Y(3940)$~\cite{Abe:2004zs,Aubert:2007vj}, 
which is also produced in $B$-decays and was found in the $J/\psi\omega$ decay channel near this respective threshold.
The mass and width of the $Y(3940)$ resonance are:
$m_{Y(3940)} = 3943 \pm 11 ({\rm stat}) \pm 13 ({\rm syst})$ MeV,
$\Gamma_{Y(3940)} = 87 \pm 22 ({\rm stat}) \pm 26 ({\rm syst})$ MeV
(Belle Collaboration~\cite{Abe:2004zs}) and
$m_{Y(3940)} = 3914.6^{+3.8}_{-3.4} ({\rm stat}) \pm 2.0 ({\rm syst})$ MeV,
$\Gamma_{Y(3940)} = 34^{+12}_{-8} ({\rm stat}) \pm 5 ({\rm syst})$
MeV ({\it BABAR}~\cite{Aubert:2007vj}).

The problematic nature of these resonances with respect to their arrangement in the constituent quark model originates in the decay patterns. A conventional pure charmonium state decays dominantly to open charm modes while hidden charm decays would be strongly suppressed~\cite{Swanson:2006st,Eichten:2007qx} due the Okubo, Zweig and Iizuka (OZI) rule. 
Under the assumption that the $Y(4140)$ is a pure $c\bar c$ state a quantitative~\cite{Liu:2009iw} estimate for the hidden charm $Y(4140)\to J/\psi \phi$ decay width results in the order of a few keV. This is in fact around several orders of magnitude smaller than current data seem to imply for the $Y(4140)\to J/\psi \phi$ decay width. Similarly, the $Y(3940)\to J/\psi \omega$ decay width is larger than 1 MeV ~\cite{Eichten:2007qx}, orders of magnitude away from the $c\bar c$ expectation. This could be signals for nonconventional structures of the $Y(3940)$ and the $Y(4140)$.
Possible alternative interpretations involve structures such as
hadronic molecules, tetraquark states or even hybrid configurations
(for recent reviews see e.g. Refs.~\cite{Swanson:2006st,Eichten:2007qx}).

The closeness of the $D^\ast\bar D^\ast$ and $D^{\ast\,+}_sD^{\ast\,-}_s$ thresholds implies a meson-meson bound state structure. But also quantitative analyses based on the possible dynamical generation of both resonances seem to support this interpretation. 
As a first follow-up to the CDF result it is suggested in~\cite{Liu:2009ei} 
that both the $Y(3940)$ and $Y(4140)$ are hadronic molecules in this case in the context of a meson-exchange 
dynamics. 
These hadron bound states can have quantum numbers $J^{\rm PC} = 0^{++}$ or 
$2^{++}$ whose constituents are the vector charm $D^\ast (D^\ast_s)$ mesons:
\eq\label{M_str} 
|Y(3940)\ra &=& \frac{1}{\sqrt{2}} \big(| D^{\ast +} D^{\ast -} \ra+ 
|D^{\ast 0} \overline{D^{\ast 0}} \ra \big)\,, \nonumber\\ 
|Y(4140)\ra &=& | D^{\ast +}_s D^{\ast -}_s \ra \,. 
\en
Earlier results based on the pion-exchange mechanism already indicated 
that the $D^\ast \overline{D^\ast}$ system can form a bound 
state~\cite{Tornqvist:1993ng}. The dominant $D^\ast \bar D^\ast$ component in the $Y(3940)$ is also confirmed in the coupled channel analysis of \cite{Molina:2009ct}. Binding in the $D^\ast_s \overline{D^\ast_s}$ 
channel can be induced by $\eta$ and $\phi$ meson exchange~\cite{Liu:2009ei}. Further on recent QCD sum rule studies~\cite{Albuquerque:2009ak,Zhang:2009st} also favor a molecular structure of the $Y(4140)$.

In the present article we report on a first quantitative prediction for
the decay rates of the observed modes $Y(3940) \to J/\psi \omega$ and
$Y(4140) \to  J/\psi \phi $ assuming the hadronic molecule structures
of Eq.~(\ref{M_str}) with quantum numbers $J^{\rm PC} = 0^{++}$ (for further details see ~\cite{Branz:2009yt}).
Results will be shown to be fully consistent with present experimental 
observations, strengthening the unusual hadronic molecule interpretation. 
Further predictions are given for the radiative two-photon
decays of these states. Finally, we also consider the alternative
$J^{\rm PC} = 2^{++}$ assignment for the $Y$ states.

\section{Theoretical background}

In order to study the decay properties of hadronic bound objects we use a method based on effective Lagrangians describing the interaction of the particles involved. Other hadron molecules such as the $a_0(980)$, $f_0(980)$, $D_{s0}^\ast(2317)$, $D_{s1}(2460)$, and $X(3872)$ have already been successfully described and studied with this method (see e.g.~\cite{Branz:2008ha,Faessler:2007us}). The composite (molecular) structure of 
the $Y(3940)$ and $Y(4140)$ states is defined by the compositeness condition 
$Z=0$~\cite{Weinberg:1962hj,Salam:1962ap} which implies that the renormalization constant 
of the hadron wave function 
is set equal to zero. Therefore the hadron exists solely as a bound state of its 
constituents. As a consequence decay processes of hadron molecules proceed in leading order via meson loops of its constituents.

For the observed $Y(3940)$ and $Y(4140)$ states  
we adopt the convention that the spin and parity 
quantum numbers of both states are $J^{\rm PC} = 0^{++}$. 
Presently the $J^{\rm P}$ quantum numbers 
are not unambiguously determined yet in experiment except for $C=+$. 
For example, the $Y(3940)$ is also discussed as a $J^{\rm PC} = 1^{++}$ 
charmonium candidate~\cite{Eichten:2007qx}, but $0^{++}$ is not ruled out. 
Their masses are expressed in terms of the binding energy
$\epsilon_Y$ as 
$m_{Y(3940)} = 2 m_{D^\ast} - \epsilon_{Y(3940)}$ and  
$m_{Y(4140)} = 2 m_{D^\ast_s} - \epsilon_{Y(4140)}$, 
where $m_{D^\ast} \equiv m_{D^{\ast \, +}} = 2010.27$ MeV and 
$m_{D^\ast_s} = m_{D^{\ast \, +}_s} = 2112.3$ MeV are the masses 
of the constituent mesons. Since the observed masses are
relatively far from the corresponding 
thresholds we do not include isospin-breaking effects that is we suppose that charged and neutral nonstrange $D^\ast$ mesons 
have the same masses. 
Following Ref.~\cite{Liu:2009ei} we consider the $Y(3940)$ 
meson as a superposition of the molecular $D^{\ast +}D^{\ast -}$ and 
$D^{\ast 0}D^{\ast 0}$ configurations, while the $Y(4140)$ is a bound state of 
$D^{\ast +}_s$ and $D^{\ast -}_s$ mesons (see Eq.~(\ref{M_str})). For the sake of simplicity we introduce the short notations $Y(3940)\equiv Y_1$ and $Y(4140)\equiv Y_2$.
The coupling of the scalar molecular states $Y_i$ to their constituents is
expressed by the phenomenological Lagrangians 
\eq\label{LY1} 
{\cal L}_{Y_1} &=& \frac{g_{Y_1}}{\sqrt{2}}\, Y_1(x) \int d^4 y\; \Phi(y^2) \, \big\{{D^{\ast 0}}^\mu\big(x+\frac{y}{2}\big)\,\overline{D^{\ast 0}}_\mu\big(x-\frac{y}{2}\big)+{D^{\ast +}}^\mu\big(x+\frac{y}{2}\big)\,D^{\ast -}_\mu\big(x-\frac{y}{2}\big)\big\}\,,\nonumber\\
{\cal L}_{Y_2} &=& g_{Y_2}\,Y_2(x) \int d^4 y\; \Phi(y^2) \, {D^{\ast +}}^\mu\big(x+\frac{y}{2}\big)\,D^{\ast -}_\mu\big(x-\frac{y}{2}\big)\,,
\en 
where $g_Y$ is the respective coupling constant.
The interaction Lagrangian is expressed by the center of mass and relative coordinates $x$ and $y$. The distribution of the constituents inside the molecular states~$Y_i$ is expressed by the correlation function $\Phi(y^2)$. The Fourier transform of the correlation function appears as a form factor in our calculations. In the present analysis we have chosen the Gaussian form of $\tilde\Phi(p_E^2/\Lambda^2_Y) \doteq \exp( - p_E^2/\Lambda_Y^2)$, 
where $p_{E}$ is the Euclidean Jacobi momentum and $\Lambda_Y$
is a size parameter with a value of about 2 GeV -- a typical 
scale for the masses of the constituents of the $Y_i$ states.

The coupling constants $g_Y$ are determined by the compositeness 
condition with
$Z_{Y_i} = 1 - \Sigma_{Y_i}^\prime(m_Y^2) = 0$, where  
$\Sigma^\prime_{Y_i}(m_{Y_i}^2) = d\Sigma_{Y_i}(p^2)/dp^2|_{p^2=m_{Y_i}^2}$ 
is the derivative of the mass 
operator $\Sigma_{Y_i}$ generated by ${\cal L}_{Y_i}(x)$.  

\begin{figure}[thbp]
\includegraphics[trim=0cm 0cm 0cm 0cm,clip,scale=0.5]{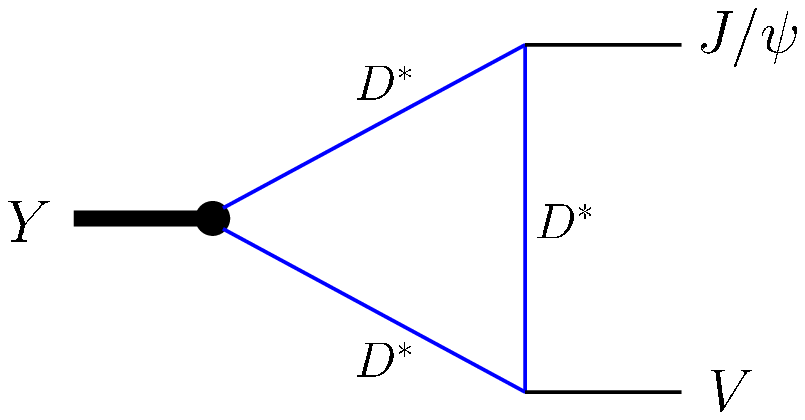}
\caption{Diagram describing the $Y \to J/\psi V$ ($V=\omega,\,\phi$) decays.}
\label{fig1}
\end{figure}

To determine the strong $Y \to J/\psi V$ and two-photon 
$Y \to \gamma\gamma$ decays we have to include the couplings 
of $D^\ast(D^{\ast}_s)$ mesons to vector mesons 
($J/\psi$, $\omega$, $\phi$) and to photons. 
The couplings of $J/\psi$, $\omega$, $\phi$ to vector 
$D^\ast(D^\ast_s)$ mesons are taken from the HHChPT 
Lagrangian~\cite{Wise:1992hn,Colangelo:2003sa}: 
\eq\label{Chiral_Lagrangian}
{\cal L}_{D^\ast D^\ast J_\psi} &=& i g_{_{D^\ast D^\ast J_\psi}}
J_\psi^\mu \Big(
  D^{\ast \dagger}_{\mu i} \,  
\!\stackrel{\leftrightarrow}{\partial}_{\nu}\!D^{\ast \nu}_i
+ D^{\ast \dagger}_{\nu i} \, 
\!\stackrel{\leftrightarrow}{\partial}^{\nu}\!\!D^{\ast}_{\mu i}
- D^{\ast \dagger\nu}_i \,  
\!\stackrel{\leftrightarrow}{\partial}_{\mu}\!D^{\ast}_{\nu i} \Big)
\,, \\
{\cal L}_{D^\ast D^\ast V} &=&  i g_{_{D^\ast D^\ast V}} V^\mu_{ij} 
D^{\ast \dagger}_{\nu i} \, 
\!\stackrel{\leftrightarrow}{\partial}_{\mu}\!D^{\ast \nu}_j 
+ 4 i f_{_{D^\ast D^\ast V}} 
(\partial^\mu V^\nu_{ij} - \partial^\nu V^\mu_{ij}) 
D^{\ast}_{\mu i} D^{\ast \dagger \nu}_j \nonumber
\en  
where $A \!\! \stackrel{\leftrightarrow}{\partial} \!\! B
\equiv A \partial B - B \partial A$, $i,j$ are flavor indices of the diagonal matrix $V_{ij} = {\rm diag}\{\omega/\sqrt{2}, \omega/\sqrt{2}, \phi\}$ containing $\omega$ and $\phi$ mesons and $D^\ast_i = (D^{\ast 0}, D^{\ast +}, D^{\ast +}_s)$ is the 
triplet of vector $D^\ast$ mesons containing light antiquarks 
$\bar u$, $\bar d$ and $\bar s$, respectively. 
The chiral couplings $g_{_{D^\ast D^\ast J_\psi}}$, 
$g_{_{D^\ast D^\ast V}}$ and $f_{_{D^\ast D^\ast V}}$ 
are taken from~\cite{Wise:1992hn,Colangelo:2003sa}
The leading-order process
relevant for the strong decays $Y(3940) \to J/\psi \omega$ and 
$Y(4140) \to J/\psi \phi$ is the diagram of Figure \ref{fig1}
involving the vector mesons $D^\ast$ or $D^\ast_s$ in the loop. Diagrammatically it already becomes obvious that the meson loop technique provides an opportunity to circumvent the OZI-suppression at lowest order in contrast to corresponding diagrams in the charmonium picture.

\begin{figure}[thbp] 
\includegraphics[trim=0cm 0cm 0cm 0cm,clip,scale=0.45]{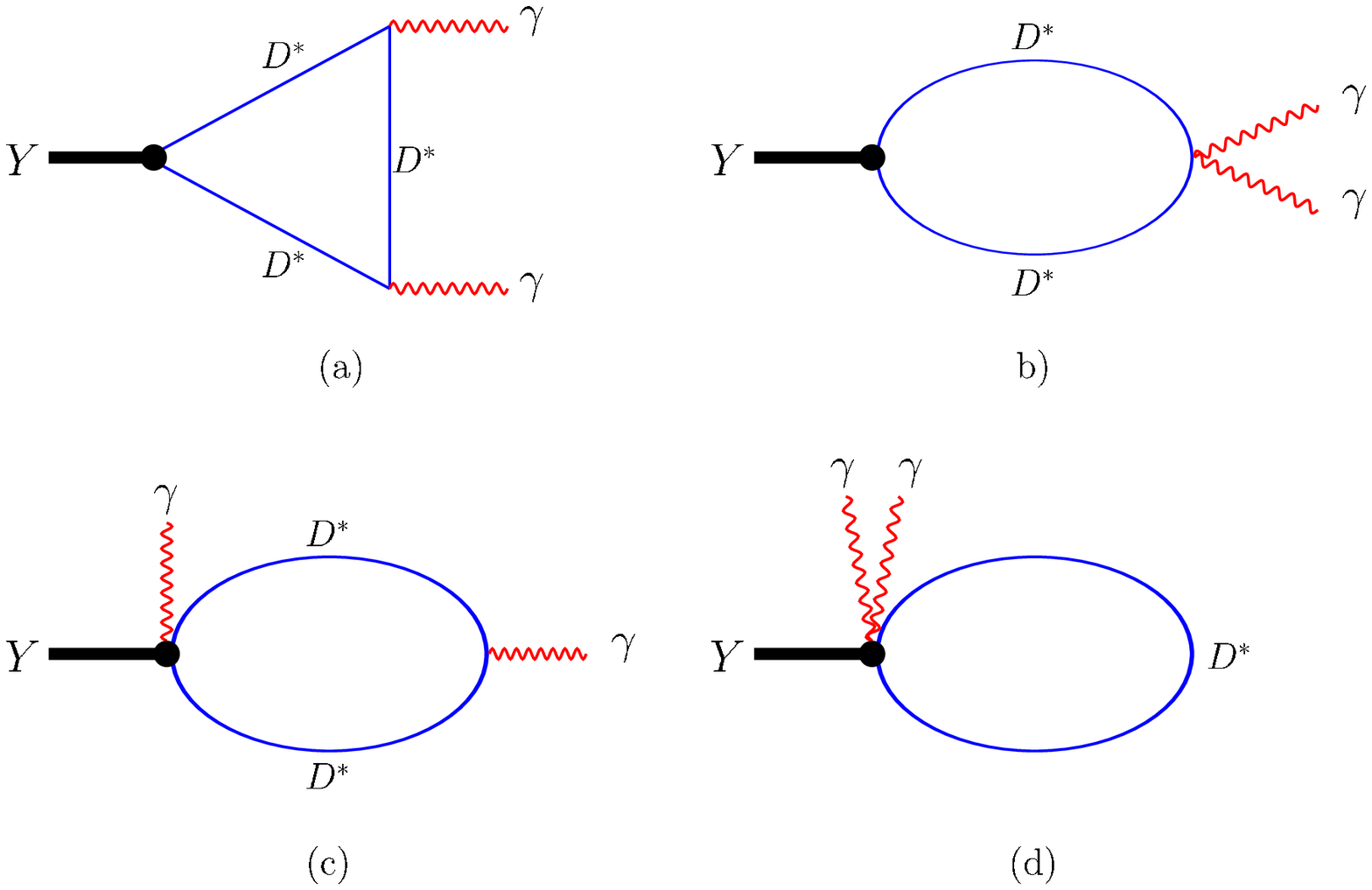}
\caption{Diagrams contributing to the 
$Y \to \gamma\gamma$ decay.}
\label{fig2}
\end{figure}

The coupling of the charged $D^{\ast \pm}(D^{\ast \pm}_s)$ mesons 
to photons is generated by minimal substitution in the free Lagrangian 
of these mesons. The corresponding electromagnetic Lagrangian is given by 
\eq 
{\cal L}_{\rm em} &=& e A_\alpha \Big( 
   g^{\alpha\nu} D^{\ast -}_\mu i\partial^\mu D^{\ast +}_\nu 
-  g^{\mu\nu} D^{\ast -}_\mu i\partial^\alpha D^{\ast +}_\nu 
+ {\rm H.c} \Big) \nonumber\\  
&+& e^2 D^{\ast -}_\mu D^{\ast +}_\nu 
\Big( A^\mu A^\nu - g^{\mu\nu} A^\alpha A_\alpha \Big)\,.
\en 
which diagrammatically leads to the two relevant graphs displayed
in Figures 2(a) and 2(b). Since the strong interaction Lagrangian of Eq. \ref{LY1} is nonlocal it has to be modified in the following way in order to restore full gauge invariance. According to~\cite{Mandelstam:1962mi} each charged constituent meson field $H^{\pm}$ in ${\cal L}_Y$ 
is multiplied by the gauge field exponential $H^{\pm}(y) \to e^{\mp i e I(y,x,P)} H^{\pm}(y)$, where 
$I(x,y,P) = \int_y^x\! dz_\mu A^\mu(z)$. This gives rise to the diagrams in Figures \ref{fig2}(c) and \ref{fig2}(d) which therefore are a consequence of the nonlocality of the present method. The contribution of these additional processes is of the order of a few percent when compared to the leading 
diagram of Figure \ref{fig2}(a). However, their inclusion is necessary to guarantee full gauge invariance.

\section{Results}

In the present analysis we studied the strong
$J/\psi V$ ($V = \omega, \phi$) and radiative two-gamma decay properties
of the $Y(3940)$ and $Y(4140)$ mesons in the framework of the effective Lagrangian approach. 
Our results for the decay widths are contained in Table~\ref{tab:1}. The errors of the results reflect the uncertainty of the experimental mass values of the $Y$ states. For the masses of the $Y$ mesons we use the values extracted by 
the {\it BABAR}~\cite{Aubert:2007vj} and the CDF~\cite{Aaltonen:2009tz} 
Collaborations.
\begin{table}[thbp]
\renewcommand{\arraystretch}{1.5}
     \setlength{\tabcolsep}{0.3cm}
     \centering
\begin{tabular}{lcc}
\hline
\tablehead{1}{l}{b}{Quantity}& \tablehead{1}{c}{b}{$\mathbf{Y(3940)}$} & \tablehead{1}{c}{b}{$\mathbf{Y(4140)}$} \\ 
\hline 
$\Gamma(Y \to J/\psi V)$, MeV       & 5.47 $\pm$ 0.34 & 3.26 $\pm$ 0.21 \\
$\Gamma(Y \to \gamma\gamma)$, keV   & 0.33 $\pm$ 0.01 & 0.63 $\pm$ 0.01 \\\hline
\end{tabular}
\caption{Decay properties of $Y(3940)$ and $Y(4140)$ for $J^{PC}=0^{++}$.}  
\label{tab:1}
\end{table} 

The predictions for the couplings $g_Y$ of the $Y$ states to their 
meson constituents are consistent with a trivial estimate using
the Weinberg formula originally derived for the
deuteron as based on the compositeness condition~\cite{Weinberg:1962hj} 
with $g_Y^W = \sqrt{32 \pi} \, m_{D^\ast}^{3/4} \, \epsilon_Y^{1/4}$.
This formula represents the leading term of an expansion in powers of the
binding energy $\epsilon$. Note that this expression can be obtained
in the local limit (i.e. the vertex function approaches the limit
$\Phi(y^2) \to \delta^4(y)$) and when the longitudinal part
$k^\mu k^\nu/m_{D^\ast}^2$ of the constituent vector meson propagator 
is neglected. The numerical results for the couplings $g_{Y(3940)}^W = 9.16$ GeV and 
$g_{Y(4140)}^W = 8.91$ GeV are rather comparable with the nonlocal results
$g_{Y(3940)} = 14.08$~GeV and $g_{Y(4140)} = 13.20$~GeV.

The predictions of $\Gamma( Y(3940) \to J/\psi \omega )=5.47$~MeV
and $\Gamma (Y(4140) \to J/\psi \phi )=3.26$~MeV for the observed
decay modes are sizable and fully consistent with the
upper limits set by present data on the total widths. 
The result for $\Gamma( Y(3940) \to J/\psi \omega )$ is also 
consistent with the lower limit of about 1 MeV~\cite{Eichten:2007qx}.   
Values of a few MeV for these decay widths naturally arise in the
hadronic molecule interpretation of the $Y(3940)$ and $Y(4140)$,
whereas in a conventional charmonium interpretation the $J/\psi V$
decays are strongly suppressed by the OZI rule~\cite{Eichten:2007qx}. In addition to the possibility of binding 
the $D^\ast \bar {D^\ast}$ and $D_s^{\ast +} D_s^{\ast -}$
systems~\cite{Liu:2009ei}, present results on the $J/\psi V$ decays give
further strong support to the interpretation of the $Y$ states as 
heavy hadron molecules.
 
Further tests of the presented scenario concern the two-photon decay
widths, which we predict to be of the order of 1 keV. 
\begin{table}[thbp]
\renewcommand{\arraystretch}{1.5}
     \setlength{\tabcolsep}{0.3cm}
     \centering
\begin{tabular}{lcc}
\hline
\tablehead{1}{l}{b}{Quantity}& \tablehead{1}{c}{b}{$\mathbf{Y(3940)}$} & \tablehead{1}{c}{b}{$\mathbf{Y(4140)}$} \\ 
\hline 
$\Gamma(Y \to J/\psi V)$, MeV       & $7.48 \pm 0.27$ & $4.41 \pm 0.16$ \\
$\Gamma(Y \to \gamma\gamma)$, keV   & $0.27 \pm 0.01$ & $0.50 \pm 0.01 $\\\hline
\end{tabular}
\caption{Decay properties of $Y(3940)$ and $Y(4140)$ for $J^{PC}=2^{++}$.}  
\label{tab:2}
\end{table}

Finally we also test the $J^{\rm PC} = 2^{++}$
assignment for the $Y$ states which is not yet ruled out experimentally. The coupling of the molecular
tensor field $Y_{\mu\nu}$ to the meson constituents is set up as
\eq\label{LY2} 
{\cal L}_{Y_1} &=& \frac{g_{Y_1}}{\sqrt{2}}\, Y_1^{\mu\nu}(x) \int d^4 y\; \Phi(y^2) \, \big\{D^{\ast 0}_\mu\big(x+\frac{y}{2}\big)\,\overline{D^{\ast 0}}_\nu\big(x-\frac{y}{2}\big)+D^{\ast +}_\mu\big(x+\frac{y}{2}\big)\,D^{\ast -}_\nu\big(x-\frac{y}{2}\big)\big\}\nonumber\\
{\cal L}_{Y_2} &=& g_{Y_2}\,Y_2^{\mu\nu}(x) \int d^4 y\; \Phi(y^2) \, D^{\ast +}_\mu\big(x+\frac{y}{2}\big)\,D^{\ast -}_\nu\big(x-\frac{y}{2}\big)
\en
Proceeding as outlined before we obtain the results in Table \ref{tab:2}. Since the results for the strong $J/\psi $ decays are quite similar to the
$0^{++}$ case, a $2^{++}$ scenario cannot be ruled out and is also
consistent within a molecular interpretation of the $Y$ states.

Quite recently the BELLE Collaboration~\cite{:2009vs} searched for the $Y(4140)$ in the two-photon process $\gamma\gamma\to J/\psi\phi$, however without evidence so far. In this framework they tested our prediction for the $Y(4140)\to \gamma\gamma$ decay width by experiment and obtained $\Gamma_{\gamma\gamma}(Y (4140)){\cal B}(Y (4140)\to  J/\psi\phi) < 40$ eV~\cite{:2009vs} for $J^P=0^+$ which results in a much smaller upper bound for the two-photon widths of about 0.2 keV. This finding is presently in conflict with a possible molecular interpretation of the $Y(4140)$.

The search for charmonium-like states in the process $\gamma\gamma\to J/\psi\omega$ by the BELLE Collaboration~\cite{uehara} resulted in an enhancement with mass $M=3915\pm3\pm2$ MeV denoted by $X(3915)$. According to its mass and width it is a possible candidate for the $Y(3940)$. The observed $2\gamma$ decay properties also support the $D^\ast \bar D^\ast$ bound state interpretation since the measured quantity, the product of the $2\gamma$ width and the branching ratio, $\Gamma_{\gamma\gamma}(X(3915)){\cal B}(X(3915)\to\omega J/\psi)=61\pm17\pm8$ eV ($J^P=0^+$) is of the same order of magnitude as the present results in Table 1.

\section{Conclusions}

The properties of the hidden-charm and two-photon decay modes of the charmonium-like mesons $Y(3940)$ and $Y(4140)$ are discussed under the assumption that both are bound states of charm mesons. The method used in the present analysis constitutes a consistent field theoretical tool for hadronic bound states and fulfills full gauge invariance. In case of the hidden charm decays $Y(3940)\to J/\psi \omega$ and $Y(4140)\to J/\psi \phi$ the sizable results for the widths obtained in the effective Lagrangian approach are in good agreement with available data and clearly support the hadron molecule assignment. A $c\bar c$ scenario is disfavored since, in contrast to the hadron molecule, the estimated widths is negligible in this case. 

In addition the two-photon decays can provide a sensitive test for the meson structure as well. In case of the $2\gamma$ width of the $Y(4140)$ available data predict a much smaller width. Note that if the present experimental upper limit for the $2\gamma$ decay width is even lower it will be very hard to explain this state as a hadronic molecule. In contrast the $Y(3940)\to \gamma\gamma$ decay width together with the sizable $J/\psi\omega$ branching ratio obtained within the $D^\ast\bar D^\ast$ bound state interpretation are comparable to experimental observations. 

However, more and particularly precise experimental information on these $Y$ states is necessary.
A full interpretation of the $Y(3940)$ and $Y(4140)$ states requires 
 an experimental determination of the $J^{\rm PC}$ quantum numbers, a consistent and hopefully converging study of binding mechanisms in 
the $D^\ast_{(s)} \overline{D^\ast_{(s)}}$ systems and finally a full understanding of the open charm decay
modes, such as $D\bar D$, $D \bar D^\ast$, $D \bar D^\ast \gamma$, etc.,
which are also naturally fed in a charmonium picture. 


\begin{theacknowledgments}
T.B. would like to thank the Organizing Committee of the HADRON 2009 for the financial support.
This work was supported by the DFG under Contract No. FA67/31-1,
No. FA67/31-2, and No. GRK683. This research is also part of the European
Community-Research Infrastructure Integrating Activity
``Study of Strongly Interacting Matter'' (acronym HadronPhysics2,
Grant Agreement No. 227431), Russian President grant
``Scientific Schools''  No. 3400.2010.2, Russian Science and
Innovations Federal Agency contract No. 02.740.11.0238.
\end{theacknowledgments}



\bibliographystyle{aipproc}   

\bibliography{refs}

\IfFileExists{\jobname.bbl}{}
 {\typeout{}
  \typeout{******************************************}
  \typeout{** Please run "bibtex \jobname" to obtain}
  \typeout{** the bibliography and then re-run LaTeX}
  \typeout{** twice to fix the references!}
  \typeout{******************************************}
  \typeout{}
 }

\end{document}